# Engineering of Atomic-Scale Flexoelectricity at Grain Boundaries

Mei Wu[1,2#], Xiaowei Zhang[1#], Xiaomei Li[3], Ke Qu[2], Yuanwei Sun[1,2], Bo Han[1,2], Ruixue Zhu[1,2], Jingmin Zhang[2], Kaihui Liu[4,5], Xuedong Bai[3], Xinzheng Li[4,5,6*] and Peng Gao[1,2,6,7*]

[1]*International Center for Quantum Materials, School of Physics, Peking University, Beijing, 100871, China*

[2]*Electron Microscopy Laboratory, School of Physics, Peking University, Beijing, 100871, China*

[3]*Beijing National Laboratory for Condensed Matter Physics and Institute of Physics, Chinese Academy of Sciences, Beijing 100190, China*

[4]*State Key Laboratory for Mesoscopic Physics, School of Physics, Peking University, Beijing 100871, China*

[5]*Collaborative Innovation Centre of Quantum Matter, Beijing 100871, China*

[6]*Frontiers Science Center for Nano-optoelectronics, School of Physics, Peking University, Beijing 100871, China*

[7]*Interdisciplinary Institute of Light-Element Quantum Materials and Research Center for Light-Element Advanced Materials, Peking University, Beijing 100871, China*

[#] These authors contributed equally to the work.

* Corresponding author. Email: p-gao@pku.edu.cn; xzli@pku.edu.cn.

**Abstract**

Flexoelectricity is a type of ubiquitous and prominent electromechanical coupling, pertaining to the response of electrical polarization to mechanical strain gradients while not restricted to the symmetry of materials. However, large elastic deformation in most solids is usually difficult to achieve and the strain gradient at minuscule is challenging to control. Here we exploit the exotic structural inhomogeneity of grain boundary to achieve a huge strain gradient (~ 1.2 nm$^{-1}$) within 3 ~ 4 unit-cells, and thus obtain atomic-scale flexoelectric polarization up to ~ 38 μC/cm$^2$ at a 24º LaAlO$_3$ grain boundary. The nanoscale flexoelectricity also modifies the electrical activity of grain boundaries. Moreover, we prove that it is a general and feasible way to form large strain gradients at atomic scale by altering the misorientation angles of grain boundaries in different dielectric materials. Thus, engineering of grain boundaries provides an effective pathway to achieve tunable flexoelectricity and broadens the electromechanical functionalities of non-piezoelectric materials.

Flexoelectricity is an emergent phenomenon that describes the electric polarization controlled by inhomogeneous deformation, i.e. strain gradients, with the broken inversion symmetry [1,2]. The flexoelectric effects have two basic features. One is its inherently universal property, which allows the appearance of net polarization from strain gradients for every dielectric material, breaking the limits of piezoelectricity that merely exits in piezoelectric material of 20 non-centrosymmetric groups [1,3]. Second, strain gradients are associated with the size scale of the system and increase with the reduction of the spatial scale [4,5]. Therefore, flexoelectric effects become pronounced at nanoscale and have potential to mediate local polar switching [6,7], piezoelectric performance [8,9], optoelectronic characteristic [10], and magnetic properties [11,12]. Particularly, the dielectric oxides have large flexoelectric coefficients and thus hold broad promising potential applications for the future nanosized devices [7,13,14]. However, it is very difficult to bend these brittle ceramic oxide materials to realize appropriate and controllable strain gradient, hindering the achievement of a high flexoelectric response.

Grain boundaries (GBs), which consist of distinct periodic arrangements of structural units, can exhibit novel properties that are not presented in the intrinsic bulk crystal [15-18]. Indeed, for a well-bonded symmetry GB, its atomic configuration of the GB core is a natural trapezoid shape with significant inhomogeneous distribution of strain in space. Such disrupted atomic bonding is usually confined at a few unit cells around the GB core, which is expected to introduce large strain gradients and resultantly conspicuous polarization at metric volumes. In addition, it's predictable and practicable to tailor the trapezoidal shape simply through changing the orientations of two adjacent grains, resulting in an adjustable strain gradient as well as flexoelectric polarization. As a result, GBs may serve as a unique active element to realize remarkable and tunable nanoscale flexoelectricity. Whereas in practice, it is challenging to validate this speculation as the GBs are in atomic size and have complex structure while the conventional macroscopic characterization and measurement only provide the collective information about the material that may be influenced by other irrelevant factors such as surface effect [19,20].

Here, we verify this speculation and develop a general strategy to generate atomic-scale flexoelectric polarization via GB engineering. From a 24º tilt LaAlO$_3$ (LAO) GB, we directly visualize the atomic arrangements by advanced atomically resolved scanning transmission electron microscopy (STEM) and spectroscopy techniques and find huge strain gradients (~1.2 nm$^{-1}$) and remarkable flexoelectric effect inducing atomic displacement within 3 ~ 4 unit cells around the GB. The first-principles density-functional theory (DFT) calculations suggest that in such a confined GB region a large local polarization (~38 μC/cm$^2$) exists and a "head-to-head" polarization configuration forms. The electron energy loss spectroscopy (EELS) illustrates stronger hybridization in La-O and Al-O interaction in the GB, facilitating the flexoelectric formation of the nanostructure. Furthermore, when extrapolated to SrTiO$_3$(STO) GB, huge strain gradients also exist in the 22.6º GB and larger strain gradients are observed in the 36.8º GB, exhibiting the alteration in misorientation angle and material for tunable flexoelectricity. The presence nanoscale flexoelectricity modifies the electrical activities and is expected to play considerable roles in the transport properties in electroceramics. Our work proves the generality and viability for atomic-scale flexoelectricity tuned by GBs, which may find potential applications in nanoelectronics and nano-electromechanical systems.

Fig. 1(a) shows the distinctive geometry of the GB in which large strain gradients are achievable. Atoms in a symmetry tilt GB are confined by the perfect bulk matrix, subsequently forming the periodic arrangements of trapezoidal structural units with large structural deformation as highlighted in Fig. 1(a). Generally, considerable flexoelectricity can only be obtained from significantly bent crystal [2]. In contrast, such an ideal atomic-sized trapezoid in the GB core is expected to generate a huge strain gradient and introduce a large dipole moment within a single unit cell. Based on the geometric relationship on the trapezoid-shaped unit (see Supplemental Material for details), we obtain the strain gradients in the GB core for different misorientation angles of GB (Fig. 1(b)).

Now we verify the possible presence of flexoelectricity at such a structure by investigating a 24º tilt LAO GB. LAO has no known ferroelectricity [21] and is an insulating oxide typically used as a gate dielectric. The LAO bicrystal was fabricated by the thermal diffusion bonding [22,23] and the correlative experimental details are presented in Supplemental Material. In order to pinpoint the atomic configurations and chemistries of that inhomogeneity, we employ the various advanced STEM techniques. Fig. 2(a) depicts a high angle annular dark field (HAADF) image (colored for clarity) viewed along the [001] direction, where structural units are periodically distributed along the [1 $\bar{5}$ 0] projection highlighted by the dotted polygon. The brighter atom columns correspond to La columns and the other type is Al, while the O columns is invisible due to the Z-contrast sensitivity of HAADF. We therefore acquire integrated differential phase contrast (iDPC) image (Fig. 2(b)), from which both cations and oxygen are directly visible [24]. Combined with atomic resolved energy dispersive X-ray spectroscopy (EDS) that can further check the possible chemical mixing in the GB (Fig. 2(c)), we ultimately determine the atomic arrangements in the buried GB, as schematically shown in Fig. 2(d). From this iDPC image, the oxygen octahedron distortions are recognizable at a few angstroms wide around the GB core. The trapezoid-shaped units in the core structure of GB impose large strain gradients and lead to substantial distortion and shifts of AlO columns, intuitively visible with the naked eye.

To better characterize the flexoelectric effect at the GB, a map of atomic displacements is extracted based on the iDPC image illustrated in Fig. 3(a), wherein the overlaid vectors represent the magnitude and direction of the offsets of AlO column with respect to the center of their surrounding four La columns. Such a cationic displacement measurement can minimize the tilt effect [25] which is evitable for GB with huge structural inhomogeneity. A stabilized "head-to-head" configuration of polar vector distribution emerges around the GB. To further evaluate the flexoelectric performance of the GB, we meanwhile quantitatively map the strain and strain gradient based on the cation sublattices from the iDPC image (Figs. 3(b) to 3(e)). The strength of strain parallel to the GB plane ($e_{xx}$) are small, while the strain gradients from $e_{zz}$ along the horizontal ($d(e_{zz})/dz$ in Fig. 3(d)) and vertical ($d(e_{zz})/dx$ in Fig. 3(e)) direction are noticeable within 1-unit cell away from the structural unit. Such a highly nonuniform strain distribution around the GB core is also verified by the geometric phase analysis (GPA) in Fig. S1 of the Supplemental Material. In particular, the unique core structure of GB [marked by red trapezoid in Fig. 3(a) and Fig. 4(a)] undergoes drastic fluctuation of strain gradients up to ~1.2 nm$^{-1}$ [$d(e_{zz})/dx$] producing significant polar displacement (~81 pm), whereas the strain gradients $d(e_{zz})/dz$ are negligible due to the symmetric geometry.

Based on the DFT calculation in Fig. 4(a), the maximum polarization of such trapezoidal-shaped unit is evaluated as ~38 $\mu C/cm^2$. The longitudinal flexoelectric coefficient $f_{11}$ of LAO then can be estimated at ~0.3 nC/m, manifesting reasonable agreement with previous work obtained from theoretical calculation of similar materials [20]. Despite of small difference due to the complex environment of the GB core structure, the overall atomic structure in the GB presented by DFT calculations is in reasonable agreement with experimental results as illustrated in Fig. S2 of the Supplemental Material. Thus the simple consideration of the off centering between cationic columns delivers the main features of polar vector distribution as shown in Fig. 3(a), excepting the red polyhedron in which the calculation suggests misalignment of O and Al in the AlO column but the experimental data cannot tell, for the reason that its contrast is likely dominated by the Al while the O is too weak to be visible. Interestingly, the polar vectors around the GB are directed towards the GB core, forming a "head-to-head" configuration with averaged ~10 μC/cm$^2$ from each side, analogous to the charged domain wall in ferroelectrics. Such a scenario certainly plays an important role in determining the electrical activities of GB in these electroceramics.

Therefore, we perform EELS of the oxygen $K$-edge and the aluminum $L$-edge to investigate the electronic structures of the flexoelectric phase in the GB. The oxygen $K$-edge (Fig. 4(b)) illustrates similar peaks at 536.5 eV in the bulk and GB core which is related to La 5d–O 2p hybridized states [26], whereas a feature around 533.5 eV (peak A) only appears in the GB core, suggesting the stronger La-O interaction [27]. Besides, feature B only appears in the GB corresponding to O 2p states hybridized with Al 3p states [27], demonstrating more hybridization between Al and O states in the GB core. Such spectral characteristics in the GB core account for the peculiar enhancement of polarization from flexoelectricity [28]. On the other hand, Al $L$-edge generally exhibits fingerprint to signify the change in Al coordination symmetry and reflect the unoccupied conducting band minimum (CBM) [29-31]. As shown in Fig. 4(c), the decrease of threshold energy on Al $L_3$ edge highlighted by the arrow is about 0.6 eV, illustrating distorted octahedral symmetries in the GB area [32], which are also reflected in the difference of fine structure on Al $L_1$ edge (see Fig. S3(a) in the Supplemental Material) [33] and substantial reduction of $L_3/L_2$ ratio depicted in Supplemental Material, Fig. S3(b). Furthermore, due to the fact that Al $L_{2,3}$ edges originate from the electron excitations from the Al-2p to unoccupied Al-3s state, while the Al-3s orbital dominates CBM [31], the lower onset energy also manifests the band gap reduction in the GB, which is also confirmed from the optical band gap measurement shown in Fig. 4(d) and the DFT calculations of the band structure (see Fig. S4 in the Supplemental Material).

To validate the generality of engineering flexoelectricity in the GB, we then investigate SrTiO$_3$ (STO) GB to present the universal structural information associated with adjustable flexoelectricity. As shown in Fig. S5 of Supplemental Material, the maximum strain gradient in 22.6º STO GB is evaluated as ~1.5 nm$^{-1}$ of d($e_{zz}$)/d$x$, resulting in cationic atomic shifts up to ~ 60 pm. The strain gradients are more pronounced in the 36.8º STO GB core [~1.9 nm$^{-1}$ of d($e_{zz}$)/d$x$], growing giant atomic shifts up to ~119 pm (see Fig. S6 in the Supplemental Material).

We would like to point out that such enhancement of displacements might not only be caused by flexoelectricity but also partly comes from the nonstoichiometry which are usually strongly coupled with each other [34] and difficult to decouple. Nevertheless, the measured strain gradients in the core of LAO and STO by experiments [depicted in Fig. 1(b) and highlight by colored dots], show good consistence with the estimated geometric relation, indicating a wide range of strain gradients can be obtained by control of the GB angle. Besides the turnability and universality, GB engineering has its advantages to generate nanoscale flexoelectricity compared to other strategies. For example, although great tensile strain exists in the center of cracks in STO [35], the symmetry stretching prevents flexoelectric polarity response due to its small strain gradients. In contrast, GB restricts the large strain gradients extending within only a few unit cells, serving as an ideal platform to deterministically triggering nanoscale flexoelectricity.

On the other hand, the polarized GB core further influences the electrostatic potential and the electronic density in the GB, modulating the electrical activities. Taking LAO as an example, a stronger hybridization state is in the GB core, favoring and promoting the formation of the ferroelectric phase. The decrease of Al $L_3$ onset energy indicates the disordered configuration complexity which may allow weak p-d hybridized states that are forbidden in perfect $AlO_6$ octahedral configuration, possibly causing a small amount of defective states between valence band maximum (VBM) states and CBM states. The shifts of band gap towards lower energy via linear fitting imply a small band gap reduction (~ 0.4 eV). Such intriguing effects decorate the trivial bulk properties, yielding potential new applications in the future nanoelectronics for GB.

In fact, although it is well-known that the GBs play critical roles in the properties of oxide ceramics, the underlying mechanism is far from clear due to the complexity. Among of all the proposed mechanisms, the elemental segregation induced local non-stoichiometricity [36,37] is widely believed to be the dominated one, which changes the local charge conditions and further influences on the transport properties [38-40]. In this study, the observed universal nanoscale flexoelectricity at GBs in these oxide ceramics provides a new scenario to understand the electrical activities, i.e., the flexoelectric dipoles carrying 'internal' polarization bound charge requires to be screened by the 'external' charges otherwise not stably exists [41], inducing charge (e.g. oxygen vacancy, hole, electron etc.) redistribution and thus being expected to influence on the electrical activities in a similar manner with non-stoichiometricity. In this sense, the role played by the huge strain gradients induced flexoelectricity is as important as the commonly believed elemental segregation for GBs.

In summary, we propose a strategy to enhance the flexoelectricity through GB engineering, where tunable and large strain gradients can be achieved at an atomic scale. By advanced STEM technique and theoretical DFT calculations, we determine the atomic structure of the exotic trapezoidal-like structure in the GBs, evaluate the strain gradients, quantify the atomic-scale flexoelectricity, reveal its electronic structure, and demonstrate the generality. Strong covalent bonding and reduction of band gap exist in the LAO GB core, which stems from symmetry-breaking effect and variations of electronic distributions perturbed by such localized polarized area within 3 ~ 4 unit cells. The universality among all dielectric materials and the adjustable dependence on GB tilt angle of this strategy provides a new effective pathway to obtain significant flexoelectricity for nanoelectronics and next-generation atomic-sized electromechanical transducers. The presence of considerable flexoelectricity in the GB also provides new insights into understanding the electrical activities of GBs in electroceramics.

This work was supported by the National Key R&D Program of China (2019YFA0708200, 2016YFA0300804, 2020B010189001), National Equipment Program of China (ZDYZ2015-1), National Natural Science Foundation of China (11974023, 52021006), and "2011 Program" Peking-Tsinghua-IOP Collaborative Innovation Center of Quantum Matter. We are grateful for the computational resources provided by the High-Performance Computing Platform of Peking University.

**Figures and captions**

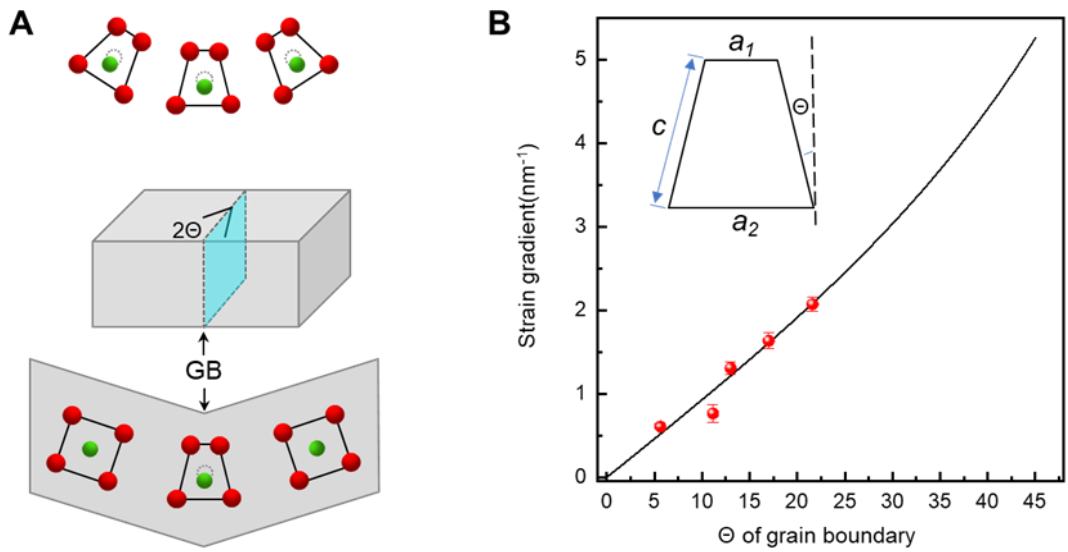

Fig. 1. Schematics for a design of large strain gradient via GB engineering. (a) Schematic display of an inhomogeneous deformation of the unit cell in the GB, producing a net dipole moment due to the flexoelectricity. (b) The expected strain gradient for GB with different tilt angles. The dots represent experiment data from LAO (blue) and STO (red) GBs.

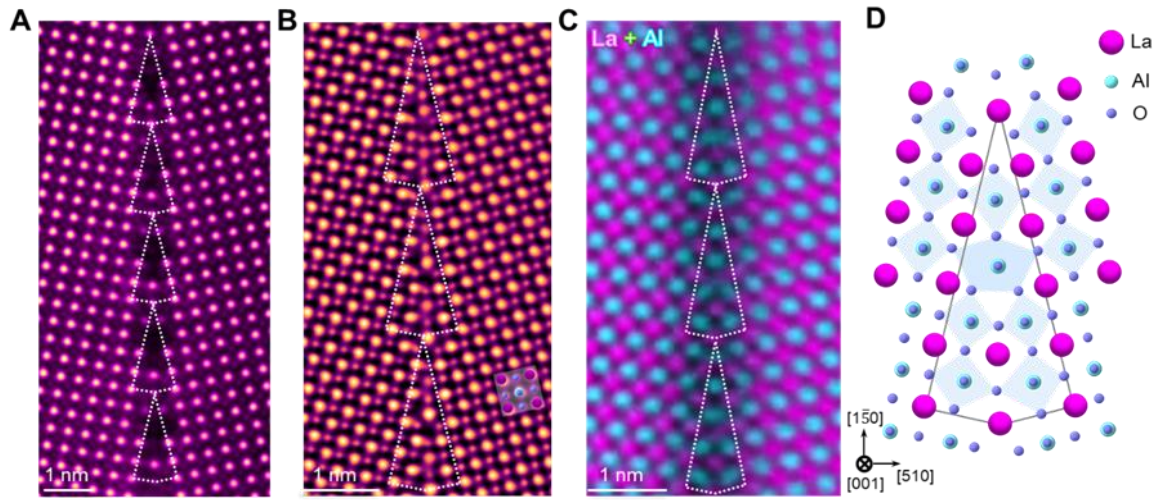

Fig. 2. Atomic arrangements of the LAO GB. (a) HAADF image. The periodic structural units are highlighted by the white polygon. (b) iDPC image depicting the oxygen and cationic arrangements. (c) Atomically resolved EDS mapping determining the cationic configuration in the GB. Red: La; Green: Al. (d) Schematic representation of the complete atomic structure of the LAO GB illustrating structural distortion at the GB. Red: La; Green: Al; Blue: O.

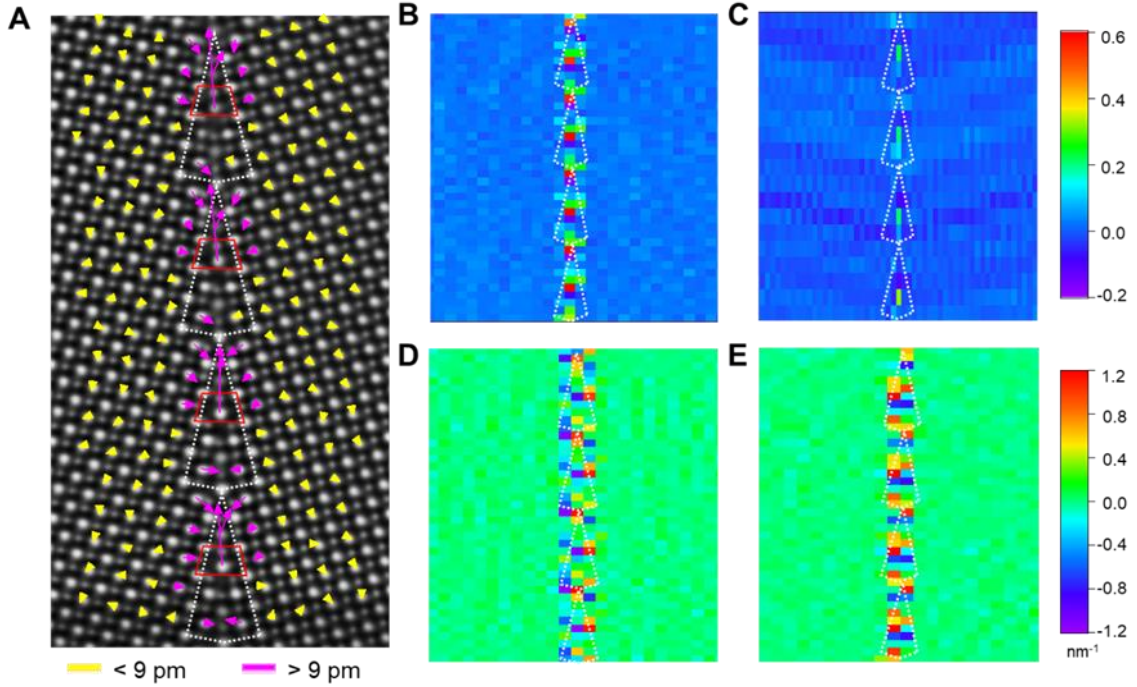

Fig. 3. Atomic-scale displacement and strain gradient in a LAO GB. (a) Off-center displacement vector map between the La and AlO columns from the iDPC image. The structural units of the GB are highlighted by the white polygon. The arrows represent the magnitude and direction of polar displacements estimated by the offsets of AlO column with respect to the center of their surrounding four La columns. (b,c) The unit cell scale mapping of $e_{zz}$ (b) and $e_{xx}$ (c) corresponding to strain perpendicular to the GB plane and parallel to the GB plane respectively. (d,e) Strength of strain gradients from $e_{zz}$ (d) and (e) correspond to the horizontal [d($e_{zz}$)/d$z$] and vertical [d($e_{zz}$)/d$x$] strain gradients respectively. The structural units of the GB are highlighted by the white polygon.

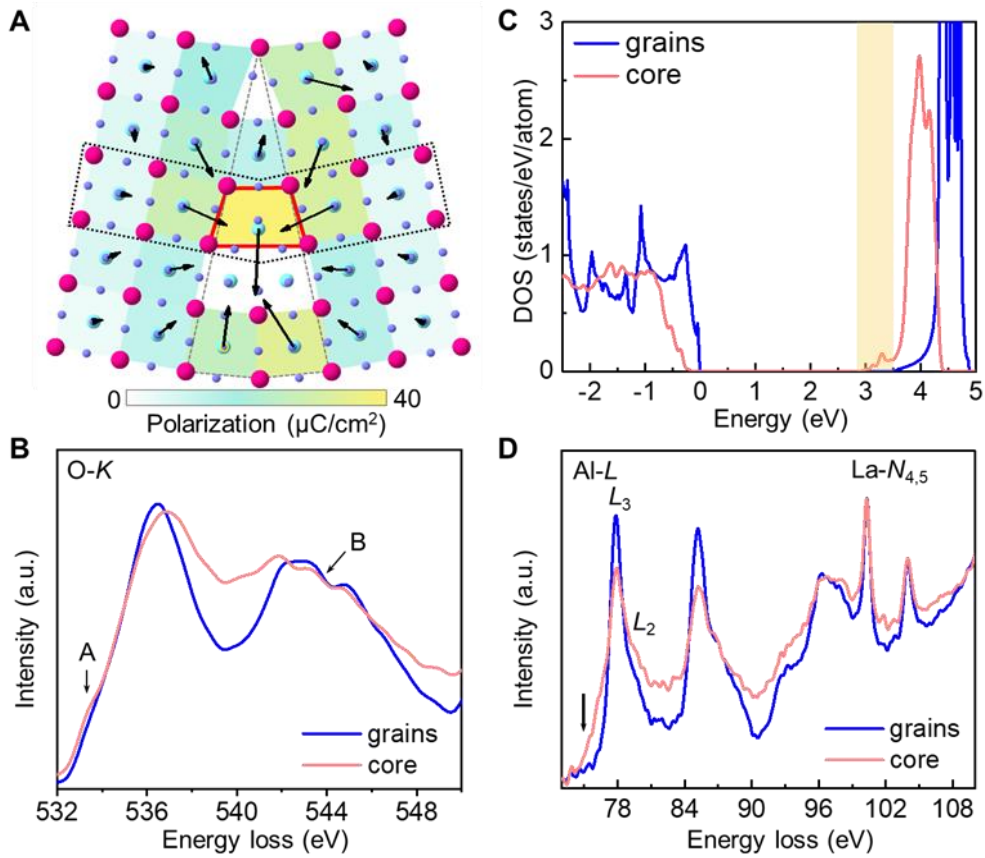

Fig. 4. Polarization and electronic structures of the LAO GB. (a) Atomic structure and polarization of the GB from DFT calculations. Vectors denote the polarization direction for each unit cell. Strength of polarization is also expressed as a color map, ranging from white (weak) to yellow (strong). In the red polyhedron, the calculation suggests misalignment of O and Al in the AlO column that is not visible to the experimental data as the contrast of AlO dominated by the Al, accounting for the deviation between polar vector and displacement vector. (b) O-$K$ edges from the GB core (red) and grains (blue) indicating stronger hybridization of La-O and Al-O in the GB. (c) Al-$L$ edges and La-$N$ edges manifesting different local Al-O configurations in the GB. (d) EELS acquired on the GB (red) and grains (blue). The extracted values of band gaps are 6.05 eV for bulk and 5.68 eV for GB core respectively.

**Methods**

**Fabrication of bicrystal.** The LAO bicrystal with 24° mistilt GB was fabricated by using the thermal diffusion bonding of two LAO single crystals at 1800 °C for 12 h in an $N_2$ atmosphere, under uniaxial stress of 2 MPa. The STO bicrystal with 36° mistilt GB was fabricated based on the thermal diffusion bonding of two STO single crystals at 1000 °C for 10 h in an $N_2$ atmosphere, under uniaxial stress of 2 MPa. The bicrystal was supplied from Hefei Ke Jing Materials Technology Co., LTD.

**Preparation of TEM samples.** The TEM specimens were first thinned by mechanical polishing and then performed by argon ion milling. The ion milling process was carried out using PIPS™ (Model 691, Gatan, Inc.) with the acceleration voltage of 3.5 kV until a hole was observed. Low voltage milling was performed with accelerating voltage of 0.3 kV to remove the surface amorphous layer and minimize the irradiation-damaged layers.

**Electron microscopy characterization and image analysis.** HAADF and iDPC images were recorded at 300 kV using an aberration-corrected FEI Titan Themis G2. The convergence semi-angle for imaging is 30 mrad, the collection semi-angles snap is 4 to 21 mrad for the iDPC imaging and 39 to 200 mrad for the HAADF. To get enough signal-noise ratio for quantitively analysis, the iDPC image was acquired at 4096×4096 pixels, with a dwell time of 500 ms/pixel and a beam current of 20 pA to avoid the beam damage.

To obtain the quantitative information on flexoelectricity, the atom positions were determined by simultaneously fitting with two-dimensional Gaussian peaks using a MatLab code. The magnitude of the strain is defined as the difference in the lattice parameters between the strained and ground states. The strain was calculated based on the La (Sr) sublattice as $strain = \frac{a-a_0}{a_0}$, where $a_0$ is the lattice parameter of free-standing LAO (STO) and $a$ is the in-plane lattice parameter perpendicular or parallel to the GB plane. The strain gradient was then the differential of strains along different direction. In Fig. 1B, we recognize side $c$ as the lattice parameter($a_0$) of the free-standing state.

The previous study [42] suggested the error of displacement measurement mainly comes from the misalignments (e.g. specimen mistilt and residual lens aberration). In our work, we conducted careful alignment to minimize the artifacts. The surrounding grains away from the GB were regarded as the reference polarization-free structure. Then the measured displacements in the reference LAO and STO is 4.1 ± 1.7 pm and 4.8 ± 1.6 pm, showing well agreement with the previous iDPC image error level (typically ~5 pm). While at the GB, the measured displacement induced by flexoelectric effect is typically an order of magnitude larger, proving the data reliability. The estimated displacements at the GB core were averaged among the periodic structural units.

**EELS characterizations and analysis.** The STEM-EELS were recorded using a Nion HERMES 200 microscope. In order to reduce beam damage, the EELS experiments were performed at 60 kV. The probe convergence semi-angle was 35 mrad and the collection semi-angle was in the range of 24.9 mrad. We acquired single 25×70 pixels EELS mapping from a region of 4.27×16 nm containing the GB. The dwell time was 150 ms/pixel, the dispersion was 0.1663 eV/ch. The data was smoothed by the Gaussian-weighted moving average filter with window length of 7 channels and variance of 1 channel. The extracted EELS spectra in Fig. 4*B* and 4*D* were spatially averaged over the regions parallel the GB plane (marked by the rectangle in *SI Appendix*, Fig. S3*B*). The EELS background was fitted and subtracted using power law $I(\Delta E) = A_0 \cdot \Delta E^{-r}$.

**First-principles calculations.** The electronic structure calculations were performed using the Vienna *Ab initio* Simulation Package (VASP) [43,44]. The projected augmented wave (PAW) [45,46] potential was used to deal with the interactions between ions and electrons. The exchange-correlation potential was treated by the Perdew-Burke-Ernzerhof [47] functional. A plane-wave cutoff of 600 eV was used. For bulk rhombohedral structure, the experimental lattice constants (a=b=c=5.360 Å, α=β=γ=60.011°) were used. Then the bulk pseudo-cubic structure was used as a block unit to build the GB structure [48]. Atomic positions were relaxed until the maximum force was less than 0.01 eV/Å. Band structures were obtained at the level of DFT. For GB, we found that the charge-neutral system is gapless. Thus, in order to make the system have a finite gap extra four electrons were added to the cell. Born effective charges were computed using the density functional perturbation theory [49-51].

The polarization calculation is based on the equation of spontaneous polarization $P_s = \frac{1}{V}\sum \delta_i Z_i$ [52], where $V$ is the volume of unit cell, $\delta$ is the displacement of atom, $i$ from its centrosymmetric position and $Z$ is the Born effective charge of atom $i$. $\delta$ is evaluated from the displacements from each column with respect to the corresponding position of polarization-free bulk crystal.